\documentclass[aps,prb,showpacs,twocolumn]{revtex4-1}
\usepackage{graphics,wrapfig,times}
\usepackage{graphicx}
\usepackage{amsmath}
\usepackage{amssymb}

\newcommand{\bvec}[1]{{\mathbf #1}}

\newcommand{\be}{\begin{equation}}
\newcommand{\ee}{\end{equation}}
\newcommand{\bea}{\begin{eqnarray}}
\newcommand{\eea}{\end{eqnarray}}

\usepackage{color}

\begin{document}

\title{Asymmetric spatial structure of zero modes for birefringent Dirac fermions}
\author{Bitan Roy$^{1,2}$, Peter M. Smith$^1$, and Malcolm P. Kennett$^1$}  
\affiliation{$^1$ Physics Department, Simon Fraser University, 8888 University Drive, Burnaby, British Columbia, V5A 1S6, Canada \\
$^2$ National High Magnetic Field Laboratory and Department of Physics, Florida State University, Tallahassee, 
Florida 32306, USA}
\date{\today}

\begin{abstract}
We study the zero energy modes that arise in an unusual vortex configuration
involving both the kinetic energy and an appropriate mass term
in a model which exhibits birefringent Dirac fermions as its low energy excitations.  
These zero modes only for an appropriate choice of relative vorticities of the mass
and kinetic energy topological defects.
We find the surprising feature that the ratio of the length scales associated with states 
centered on vortex and anti-vortex topological defects can be arbitrarily varied
but that fractionalization of quantum numbers such as charge is unaffected.  
We discuss this situation from a symmetry point of view and present numerical
results for a specific lattice model realization of this scenario.
\end{abstract}


\maketitle

\section{Introduction}

It is well known that the Dirac Hamiltonian with topologically
non-trivial mass terms allows for zero energy 
modes.\cite{JackiwRebbi,Su,JackiwRossi,Weinberg,ReadGreen,Hou,JackiwPi,Seradjeh,Seradjeh2} 
If there is a single zero energy mode, then this leads to the non-trivial
phenomenon of quantum number fractionalization, e.g. \emph{charge} as   
 has been experimentally confirmed in one dimension in polyacetylene.\cite{Su,polyacetelene}
More recently, there has been much interest in zero modes that can arise
from topological defects in systems whose low energy excitations can
be described using Dirac 
fermions.\cite{Topdef,Chamon,Nishida,ChiKen1,ChiKen2,Hosur,Cooper,Goldstein,Igorpseudospin} 
Such modes are being studied intensely due to the 
possibility of their application in quantum computation.\cite{Majorana,Akhmerov,Ivanov,Nayak}  

In all of these examples, the zero modes 
arise as a consequence of a topological defect in the 
{\it mass} term of the Dirac Hamiltonian.  We consider here
the recently introduced model of birefringent fermions
\cite{Kennett} and introduce a momentum space vortex
in addition to a standard mass vortex.  The low energy theory of
birefringent fermions consists of four component massless 
fermions with two separate
Fermi velocities $v_0 (1 \pm \beta)$ controlled by
the parameter $0 \leq \beta \leq 1$.  Writing the low energy theory in 
Dirac form, the parameter $\beta$ multiplies 
terms in the kinetic energy not present in the regular Dirac
Hamiltonian. 

Our main result is that when there is 
an appropriate vortex (anti-vortex)
in $\beta$ in addition to a mass vortex (anti-vortex) 
of the type introduced in Ref.~\onlinecite{Hou}
then the ratio of the characteristic lengthscales of the
zero mode solutions in the presence of either a vortex or an anti-vortex
may be tuned arbitrarily by $\beta$.  
Specifically, we find that finite $\beta$ breaks the symmetry present
when $\beta = 0$ between the zero mode solutions in the presence of a 
mass vortex and a mass anti-vortex.  The zero mode solution in the 
presence of a vortex becomes more extended while with an underlying 
anti-vortex it becomes more localized.
In the limit $\beta = 0$ the model we consider displays the same
physics as that discussed in Ref.~\onlinecite{Hou}.  
The type of vortex we consider here may also be
of interest in a variety of systems whose low energy excitations
can be described as Weyl fermions with multiple Fermi velocities, which have recently been
the focus of a number of publications.\cite{Bercioux,Watanabe,Lan1,Lan2,Goldman,Moessner}

This paper is structured as follows.  In Sec.~\ref{sec:model} we
recall the model of birefringent Dirac fermions and describe the vortex
that we are considering.  In Sec.~\ref{sec:zero} we find the zero
energy modes associated with this vortex, and study a tight-binding lattice
model to illustrate this physics numerically in Sec.~\ref{sec:numerics}. 
In Sec.~\ref{sec:disc}
we discuss our results and conclude.

\section{Model and Vortex}
\label{sec:model}
\begin{figure}[htb]
\includegraphics[width=5cm]{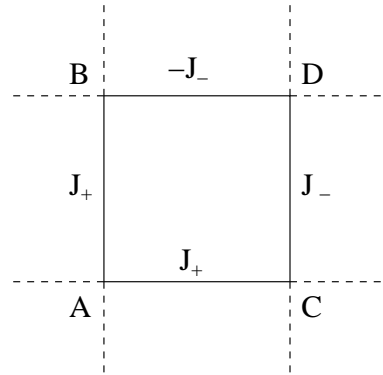}
\caption{Unit cell of tight binding model with birefringent Dirac fermions
as low energy excitations.\cite{Kennett}  Hopping parameters are 
indicated, with $J_\pm = J_0 (1 \pm \beta)$.}
\label{fig:unitcell}
\end{figure}

The model of birefringent fermions introduced in Ref.~\onlinecite{Kennett}
has the feature that there are massless fermions near the Dirac points, but with two
distinct Fermi velocities.  This model may 
be obtained as the low-energy theory associated with spinless fermions
at half filling in 
a particular tight binding model on a square lattice 
with a four site unit cell illustrated in Fig.~\ref{fig:unitcell}.

The dispersion relation reads as
$$ E_{k} = \pm J_\pm \sqrt{\cos^2 k_x + \cos^2 k_y} , $$
where $J_\pm = J_0 (1 \pm \beta)$, with $0 \leq \beta \leq 1$.  This 
dispersion leads to four equivalent Dirac points at the corners of the 
Brillouin zone: $\bvec{K}_{\pm,\pm} = 
\left(\pm\frac{\pi}{2},\pm\frac{\pi}{2}\right)$.
Labelling the four sites in the unit cell 
as $A$, $B$, $C$, and $D$ we can write the low
energy theory in the form:
\begin{eqnarray}
\label{model}
H = \sum_{\bvec{k}} \psi_k^\dagger[E_k - H_k] \psi_k ,
\end{eqnarray}
where $ \psi_k^T = (c_{Ak}, c_{Bk}, c_{Ck}, c_{Dk}),$
with $c_{Ik}$ a fermionic annihilation operator for 
a fermion with momentum $k$ which resides on sites $I = A, B, C,$ or $D$,
and (setting $2J_0 = 1$)
\begin{eqnarray}
H_k = \left[i\left(\gamma_0 \gamma_1 + i\beta\gamma_3\right)k_x
+ i\left(\gamma_0 \gamma_2 + i\beta \gamma_5\right)k_y\right].
\label{eq:birefHam}
\end{eqnarray}
We use a non-standard representation of the gamma matrices in
which $\gamma_0 = \sigma_3 \otimes \sigma_3$, $\gamma_1 = \sigma_2
\otimes I_2$, $\gamma_2 = \sigma_3 \otimes \sigma_2$,
$\gamma_3 = -\sigma_1\otimes I_2$, 
and $\gamma_5 = \gamma_0 \gamma_1 \gamma_2 \gamma_3  = -\sigma_3\otimes \sigma_1$.
The matrices $\gamma_0$, $\gamma_1$, $\gamma_2$, $\gamma_3$ and $\gamma_5$
satisfy the Clifford algebra $\gamma_\mu \gamma_\nu + \gamma_\nu
\gamma_\mu = 2\delta_{\mu\nu}$.\cite{footnote}   The representation of the
gamma matrices is four dimensional, which is
the minimal dimension for a time-reversal invariant system of spinless
Dirac fermions in two dimensions on a lattice.\cite{Herbut}  By way of 
comparison, in graphene, the minimal 
representation is constructed with \emph{two} sublattice degrees of freedom 
and \emph{two} inequivalent Dirac points.  In the present problem, on the other hand,
there are four equivalent Dirac points, but the unit cell is comprised 
of \emph{four} lattice points.   The two Dirac cones with Fermi velocities $1 \pm \beta$
suggest the problem may be written as 
a direct sum of two copies of two component massless 
Dirac fermions or Weyl fermions.
[This possibility was recently considered and it was shown that 
one cannot have a Weyl fermion in this present scenario.\cite{Herbut}] 
The realization of Weyl fermions in a 
two dimensional lattice without broken time reversal symmetry is also 
prohibited by Nielsen-Ninomiya theorem.\cite{nielsen, haldane} However in the 
limit $\beta = 1$ there are two flat bands at zero energy and a two 
component Weyl fermion. Such a possibility was previously discussed in other
 contexts\cite{dagotto,ShenApaja} -- 
nevertheless the Nielsen-Ninomiya theorem is respected there as well. 

\subsection{Topological Defect}
\label{sec:vortex}
When $\beta = 0$ in the birefringent fermion model,
there is a chiral SU(2) symmetry generated by $\gamma_3, \gamma_5,$ and 
$\gamma_{35}$, where $\gamma_{35} = i\gamma_3 \gamma_5$, but this
symmetry is broken when $\beta \neq 0$.\cite{Kennett}  If we consider the Hamiltonian 
[Eq.~(\ref{eq:birefHam})] with non-zero
$\beta$ and introduce a vortex such that $\beta \to 0$ in the centre, then 
chiral symmetry is restored in the centre of the vortex.  Taking $\beta$
to be solely a function of the radial co-ordinate $r$, that
vanishes at the origin and has a constant limit
as $r \to \infty$, then an appropriate Hamiltonian for an anti-vortex in 
$\beta$ reads as

\bea
H_\beta & = & i\gamma_0\gamma_1 (-i\partial_x) + i\gamma_0 \gamma_2 (-i\partial_y)
\nonumber \\
& & -
\beta(r)\left(\cos (v_\beta \theta) \gamma_3
 - \sin (v_\beta\theta) \gamma_5\right) (-i\partial_x) \nonumber \\
& &  - \beta(r)\left(\cos(v_\beta\theta) \gamma_5 +
 \sin(v_\beta\theta) \gamma_3\right) (-i\partial_y)
\label{eq:betavort}
\eea
The vorticity $v_\beta$ is a non-negative integer.
One may arrive at Eq.~(\ref{eq:betavort}) from Eq.~(\ref{eq:birefHam}) by 
the substitution $\beta \to \beta(r)$ and a
chiral rotation $U_c=e^{i v_\beta \theta \gamma_{35}}$ in Eq.~(\ref{eq:birefHam}).  
Without a mass term leading to a gap in the spectrum, this 
vortex in $\beta$ does not support a normalizable zero mode.

In order to obtain zero modes, we need a mass term in the 
Dirac-like Hamiltonian to open a gap. 
An appropriate term to consider is the mass term considered by 
Hou {\it et al.}\cite{Hou} in the context of graphene.  In that 
context, the term corresponds to a staggered hopping in a Kekule
pattern.  We show that this term leads to gap for birefringent 
fermions in Appendix~\ref{app:gap}.  In the square lattice problem of Ref.~\onlinecite{Kennett}, the 
equivalent term is also a staggered hopping, which leads to contribution
in the low energy Hamiltonian of
\begin{equation}
H_m = -m(r) \; \left(\cos(v_m\theta) i\gamma_0 \gamma_3 
- \sin(v_m\theta) i\gamma_0\gamma_5\right) ,
\end{equation}
corresponding to a mass anti-vortex with spatial profile given by $m(r)$ and vorticity
$v_m$, which is a non-negative integer. Note that $i\gamma_0 \gamma_3$ corresponds to staggered
hopping in the $x$ direction and $i\gamma_0\gamma_5$ corresponds
to staggered hopping in the $y$ direction. We will
hence work with the Hamiltonian $H_{m,\beta}=H_\beta \;+\;  H_m$.
In the limit $\beta=0$, 
the Hamiltonian $H_{m,\beta}$ takes the form of the massive Dirac Hamiltonian 
originally studied by Hou {\it et al}. \cite{Hou}  An 
index theorem therefore guarantees the existence of $v_m$ 
normalizable zero energy states \cite{JackiwRossi, Weinberg} and concomitant fractionalization 
of charge when $v_m=1$ and $\beta =0$.\cite{Hou}  
Here we seek to find the zero energy modes in the spectrum
of $H_{m,\beta}$ with $\beta \neq 0$ and study how the anti-vortex in $\beta$
affects the zero modes that arise from $H_m$.

For calculational convenience, we make use of a unitary transformation $U = U_3 U_2 U_1$,
where $U_1 = I_2 \oplus \sigma_2$, $U_2 = \frac{1}{\sqrt{2}}
\left(I_4 - i \sigma_1\otimes I_2\right),$ and $U_3 = \frac{1}{\sqrt{2}} 
\left(I_4 + i \sigma_2\otimes \sigma_3\right)$ to the ``graphene representation'' of the 
gamma matrices: $\gamma_0^G = I_2\otimes\sigma_3$, $\gamma_1^G = \sigma_3 
\otimes \sigma_2$, $\gamma_2^G = I_2\otimes\sigma_1$, $\gamma_3^G = 
\sigma_1\otimes\sigma_2$, and $\gamma_5^G = \sigma_2\otimes\sigma_2$.\cite{herbut-juricic-roy}
Then the transformed pieces of the Hamiltonian take the form
\begin{eqnarray}
H_\beta & = & i\gamma_0\gamma_1 (-i\partial_x) - i\gamma_0 \gamma_2 (-i\partial_y) 
\nonumber \\ & &
- \beta(r)\left(-\cos(v_\beta\theta) \gamma_3
 + \sin(v_\beta\theta) \gamma_5\right) (-i\partial_x) \nonumber \\ & & 
 - \beta(r)\left(\cos(v_\beta\theta) \gamma_5 +
  \sin(v_\beta\theta) \gamma_3\right) (-i\partial_y) ,
\end{eqnarray}
and
\begin{equation}
H_m = -m(r)\left(-\cos(v_m\theta) i\gamma_0 \gamma_3 - \sin(v_m\theta) i\gamma_0\gamma_5\right).
\label{eq:mass-term}
\end{equation} 
We assume that the spatial profiles of the defects in $\beta$ and $m$ are that as $r 
\to 0$, $\beta(r) \to 0$ and $m(r) \to 0$, and in the large $r$ limit 
(far from the core of the vortex), $\beta(r) \to \beta_0$ and $ m(r) \to
m_0$, where $\beta_0$ and $m_0$ are constants.  It might be possible to realize
such a vortex configuration experimentally by implementing a pattern of hopping 
integrals such as described in Sec.~\ref{sec:numerics} for cold atoms in 
an optical lattice\cite{Bakr,Schonbrun,Bloch,Tarruell} or through other 
synthetically constructed systems with Dirac fermionic excitations.\cite{Gomes}

\section{Zero energy modes}
\label{sec:zero}

We now search for zero energy modes that satisfy $H_{m,\beta}  \Psi =0$.  
The existence of a
unitary operator, $\gamma_0$, such that $\{H_{m,\beta},\gamma_0 \}=0$, 
ensures the spectral symmetry of the energy eigenstates of $H_{m,\beta}$. Moreover,
one may find an anti-unitary operator $M = UK$, where $U$ is unitary
and $K$ is the complex conjugation operator, which anticommutes with the 
Hamiltonian.\cite{ChiKen1} 
Noting that $\{i\gamma_0\gamma_1, \gamma_5, i\gamma_0\gamma_3\}$ are real 
 and $\{i\gamma_0\gamma_2, \gamma_3,i\gamma_0\gamma_5\}$ are imaginary 
one finds $U = -i\gamma_2 \gamma_3  = \sigma_1 \otimes \sigma_3$, in the 
``graphene representation". A zero energy mode of $H_{\beta, m}$ must also, therefore
be an eigenstate of $M$, which for the state $\Psi_0^T = (\psi_1, \psi_2, \psi_3, \psi_4)$
leads to the constraint $\psi_1 = \pm \psi_3^*$, $\psi_2 = \mp \psi_4^*$. 
The symmetry of the energy 
spectrum about zero implies that the zero energy mode should be robust
against any weak local perturbation. 

For further ease of calculation we  redefine the components of $\Psi$ as
$\Psi \to e^{\frac{i\pi}{4}\gamma_0}\Psi$
so that 
the eigenvalue equations for the zero energy mode take the form
\begin{eqnarray} 
\partial_z \psi_2 - \left(\beta(r) e^{-iv_\beta\theta} \partial_{\bar{z}} -  m(r)
e^{-iv_m\theta}\right) \psi_4 
& = & 0 ,
\label{eq:hpsizero1} \\
\partial_{\bar{z}}\psi_1 - \left(\beta(r) e^{-iv_\beta\theta} \partial_{\bar{z}} + 
 m(r) e^{-iv_m\theta}\right)
\psi_3 & = & 0 ,
\label{eq:hpsizero2} \\
\partial_{\bar{z}} \psi_4 - \left(\beta(r) e^{iv_\beta\theta} \partial_z - 
 m(r) e^{iv_m\theta}\right)\psi_2 
& = & 0 ,
\label{eq:hpsizero3} \\
\partial_z \psi_3 - \left(\beta(r) e^{iv_\beta\theta} \partial_z + 
 m(r) e^{iv_m\theta} \right)\psi_1 & = & 0 ,
\label{eq:hpsizero4}
\end{eqnarray}
where $\partial_z = \partial_x - i\partial_y = e^{-i\theta} \left(\partial_r 
-\frac{i}{r}\partial_\theta\right) $
and  $\partial_{\bar{z}} = \partial_x + i\partial_y = 
e^{i\theta} \left(\partial_r +\frac{i}{r}\partial_\theta\right)$.
$\Psi$ must be an eigenvector of $M$ as well as $H$, and choosing $\Psi$ to have
eigenvalue -1 implies $\psi_3 = -\psi_1^*$ and $\psi_4 = \psi_2^*$ (we will
discuss the choice that $M$ has eigenvalue $+1$ below).  
Under this constraint Eqs.~(\ref{eq:hpsizero1}) - (\ref{eq:hpsizero4}) 
lead to only two independent equations:
\begin{eqnarray}
\partial_z \psi_2 - \left(\beta(r) e^{-iv_\beta\theta} \partial_{\bar{z}} - 
 m(r) e^{-iv_m\theta} \right)\psi_2^* 
& = & 0 ,
\label{eq:conj1} \\
\partial_{\bar{z}} \psi_1 + \left(\beta(r) e^{-iv_\beta\theta} 
\partial_{\bar{z}} + m(r) e^{-iv_m\theta}\right)\psi_1^* & = & 0 .
\label{eq:conj2}
\end{eqnarray}
To solve these equations we make use of an ansatz introduced in a different 
context by Ghaemi and Wilczek,\cite{Ghaemi} focusing first on Eq.~(\ref{eq:conj1}).
We make the ansatz
\begin{equation}
\psi_2(r,\theta) = e^{il\theta} \phi_2(r) + e^{in\theta}\phi_4(r),
\label{eq:ghaemiansatz24}
\end{equation}
where $\phi_2$ and $\phi_4$ are real.  In order to have a consistent solution,
we must have that $v_m = v_\beta - 1$.  There is still some freedom in the 
choice of $l$ and $n$, depending on which terms are grouped together from
Eq.~(\ref{eq:conj1}) after the use of the ansatz in Eq.~(\ref{eq:ghaemiansatz24}).
The choice that allows either $l=n$ or $l \neq n$ and guarantees that the solution is 
single-valued is:
\begin{eqnarray}
\left[\partial_r + \frac{l}{r}\right]\phi_2 - \beta(r)
\left[\partial_r + \frac{n}{r}\right]\phi_4 + m(r)\phi_4 & = & 0 , 
\label{eq:phi24a}
\\
\left[\partial_r + \frac{n}{r}\right]\phi_4 - \beta(r)
\left[\partial_r + \frac{l}{r}\right]\phi_2 + m(r)\phi_2 & = & 0 ,
\label{eq:phi24b}
\end{eqnarray}
in which case we have the condition $l+n = 2 - v_\beta$. We now
consider the asymptotic behavior of the solution of this equation at large and small
$r$. In the large $r$ limit $\beta(r) \to \beta_0$ and $ m(r) \to
m_0$, and we can ignore $1/r$ terms.  This leads to the solution as $r \to \infty$
\be
 \phi_2(r) = \phi_4(r) = A \; e^{-\kappa_- r},
\label{eq:antivortex-infinity}
\ee
where $$\kappa_- = \frac{m_0}{1 - \beta_0},$$
is the inverse of the characteristic length scale for the zero mode in the 
presence of an anti-vortex
and $A$ is a normalization constant.  At small $r$, 
$\beta(r) \to 0$ and $m(r) \to 0$,
in which case 
\be
 \phi_2(r) = a_2 \; r^{-l}; \quad \quad \phi_4(r) = a_4 \; r^{-n}.
\label{antivortex-origin}
\ee
In order that solution be normalizable near the origin, we require
$l$, $n$ $\leq 0$, which implies $v_\beta \geq 2$.

We now consider the two cases $v_\beta$ odd and even separately. If $v_\beta$
is even, i.e. $v_\beta = 2p$ for some integer $p$, then to satisfy
Eqs.~(\ref{eq:phi24a}) and  (\ref{eq:phi24b}) subject to $l$, $n$ $\leq 0$ 
and $l+n = 2 - v_\beta = 2(1-p)$
requires $p \geq 1$. There will be a solution with $l=n$ and $(2p-2)$ solutions with 
$l\neq n$. As far as the original equation we were trying to solve, Eq.~(\ref{eq:conj1}), 
is concerned, solutions with $l \leftrightarrow n$
are equivalent, so there are in fact only $p-1$ solutions of Eq.~(\ref{eq:conj1})
with $l\neq n$, and a total of $p$ zero mode solutions when $v_\beta = 2p$ 
(which implies $v_m = 2p-1$).  When $v_\beta$ is odd, we can apply a similar
analysis and find that for $v_\beta = 2p +1$ (i.e. $v_m = 2p$), then there
are $p$ zero mode solutions.   Hence to have a single zero mode with $M$ eigenvalue $-1$, 
we can either have
$v_\beta = 2$, $v_m = 1$, or $v_\beta = 3$, $v_m = 2$.  

In our solution above we assumed that the eigenvalue of $M$ is $-1$.  If, 
on the other hand, we try to obtain a solution which is an eigenvector of $M$
with eigenvalue $+1$, then we find that we must have $v_\beta \geq 2$ and 
$v_m = v_\beta - 1$, as before, 
and that there is no normalizable solution when $v_\beta = 2$.  
However, there can be normalizable solutions when $v_\beta > 2$, and in particular
there is one normalizable solution when $v_\beta = 3$, so that there are a total
of $2$ normalizable zero modes (as would be expected from $v_m = 2$ by the usual
index theorem\cite{JackiwRossi,Weinberg}).
We consider the situation
with $v_\beta = 2$, $v_m = 1$ numerically in Sec.~\ref{sec:numerics}.
It should be noted that the fact that the solution we obtained from Eqs.~(\ref{eq:conj1}) and
(\ref{eq:conj2}) has eigenvalue
$-1$ for $M$ is not significant.  We could have equally well found a normalizable
solution with eigenvalue $+1$ and no normalizable solution with $-1$ when $v_\beta = 2$ 
had we made
a different redefinition of the fields before Eqs.~(\ref{eq:hpsizero1}) 
to (\ref{eq:hpsizero4}), e.g. $\Psi \to e^{-\frac{i\pi}{4}\gamma_0}\Psi$.

We can also see that the mass anti-vortex is required in order to have normalizable
solutions.  If we set $m_0 = 0$, then we have an anti-vortex defect 
in the kinetic energy alone. At large distance, the solutions 
asymptote to a constant value as $r \to \infty$
\be
 \phi_2(r) = \phi_4(r) \to A , 
\label{eq:antivortex-infinity-m0}
\ee 
leading to a non-normalizable solution. Note that additionally, when $m_0 = 0$
there is no gap in the spectrum. When $v_\beta > 0$, there are no non-zero 
normalizable solutions for $\psi_1$.

Next we consider vortex configurations in $\beta$ and $m$, corresponding to negative
integer values of $v_\beta$ and $v_m$, which will illustrate how the $\beta$ term
modifies the usual solution to the mass vortex.\cite{Hou}  In this case, 
$\psi_2$ has no normalizable solutions. Nevertheless, one can obtain 
the zero energy mode by using the ansatz
\begin{equation}
\psi_1(r,\theta) = e^{il\theta} \phi_1(r) + e^{in\theta}\phi_3(r),
\label{eq:ghaemiansatz13}
\end{equation}
and $\psi_3 = -\psi_1^*$, associated with the $-1$ eigenvalue of $M$. The coupled 
differential equations for the zero mode then read as  
\begin{eqnarray}
\left[\partial_r - \frac{l}{r}\right]\phi_1 + \beta(r)
\left[\partial_r + \frac{n}{r}\right]\phi_3 + m(r)\phi_3 & = & 0 ,
\label{eq:phi13a}
\\
\left[\partial_r - \frac{n}{r}\right]\phi_3 + \beta(r)
\left[\partial_r + \frac{l}{r}\right]\phi_1 + m(r)\phi_1 & = & 0.
\label{eq:phi13b}
\end{eqnarray}
Even though in the vicinity of the origin $\phi_1$ and $\phi_3$ behave identically to 
$\phi_2$ and $\phi_4$ respectively in Eq.~(\ref{antivortex-origin}), 
at large distances ($r\rightarrow \infty$)
\be
\phi_1(r) = \phi_3(r) = B e^{-\kappa_+ r},
\label{eq:vortex-infinity}
\ee
where 
$$ \kappa_+ = \frac{m_0}{1+\beta_0}, $$
is the inverse of the characteristic lengthscale for the zero mode in the presence of a vortex,
and $B$ is a normalization constant. The lengthscale for the decay of the zero 
mode differs in this case from the lengthscale we found for the anti-vortex
solution. The norm of the zero modes with $+1$ eigenvalue 
of $M$ in this situation grow with the system size when $v_\beta = -2$, $v_m = -1$. 

In the limit $\beta_0=0$, the characteristic length 
 scales for vortex and anti-vortex are the same,
since the Hamiltonian associated with those two distinct topological 
defects are unitarily equivalent to each other. Namely, the 
vortex Hamiltonian may be obtained from the anti-vortex one via a unitary rotation by
$\gamma_5$ in Eq.~(\ref{eq:mass-term}). However, this unitary equivalence breaks 
down for a birefringent Dirac Hamiltonian, leading to the distinct length scales 
found in Eqs.~(\ref{eq:antivortex-infinity}) and (\ref{eq:vortex-infinity}). 

It is useful to present the zero modes in terms of the original spinor 
components $\psi_k$. In the presence of an anti-vortex the zero mode reads 
as $\psi^T_0=\left(0,\psi_2,0,\psi^*_2 \right)$. 
After unitary rotation by $U=U^\dagger_1 U^\dagger_2 U^\dagger_3$ one finds 
that the amplitudes of the zero energy mode with an underlying anti-vortex
is finite only on the $B$ and $C$ sublattices. When there is an underlying vortex,
the zero modes acquire a finite expectation value on the $A$ and $D$ sublattices. 
This can also be seen by noting that both the vortex and anti-vortex Hamiltonians 
anticommute with $\gamma_0$. Therefore the zero energy subspace ${\cal H}_0$ is invariant 
under $\gamma_0$ and it acts like an \emph{identity} matrix in ${\cal H}_0$. 
Therefore all the zero energy states must be an eigenstate of 
$\gamma_0$ with eigenvalue $+1$ or $-1$. Since 
$\gamma_0 =\sigma_3 \otimes \sigma_3 \equiv {\rm Diag}(1,-1,-1,1)$, 
the amplitude of the zero modes can be finite either on the $B$ and $C$ sublattices 
(anti-vortex) or $A$ and $D$ sublattices (vortex) in accordance with our explicit 
calculation.  

\section{Numerics}
\label{sec:numerics}
\begin{figure}[htb]
\begin{center}
a) \includegraphics[width=7cm]{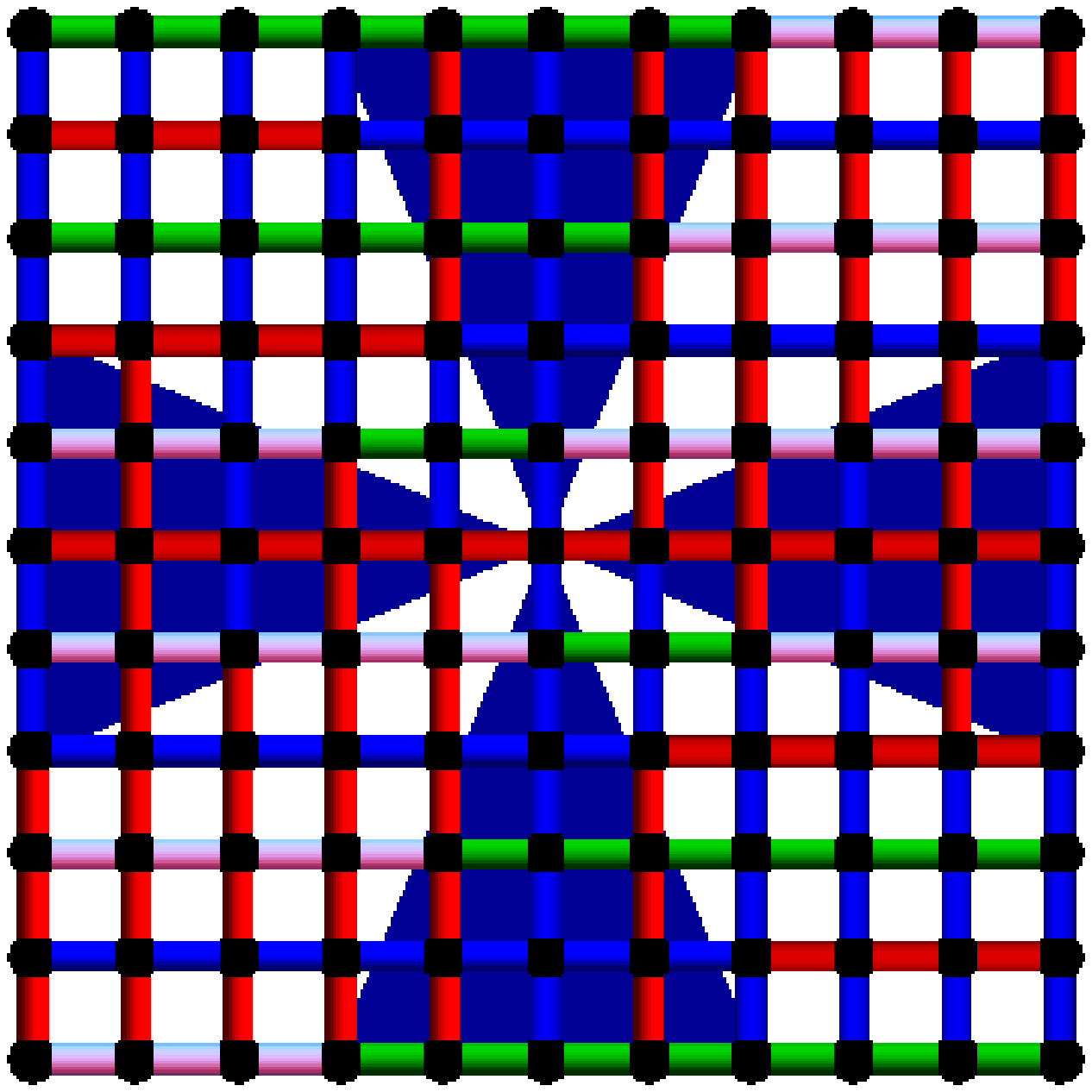} \\
b) \includegraphics[width=7cm]{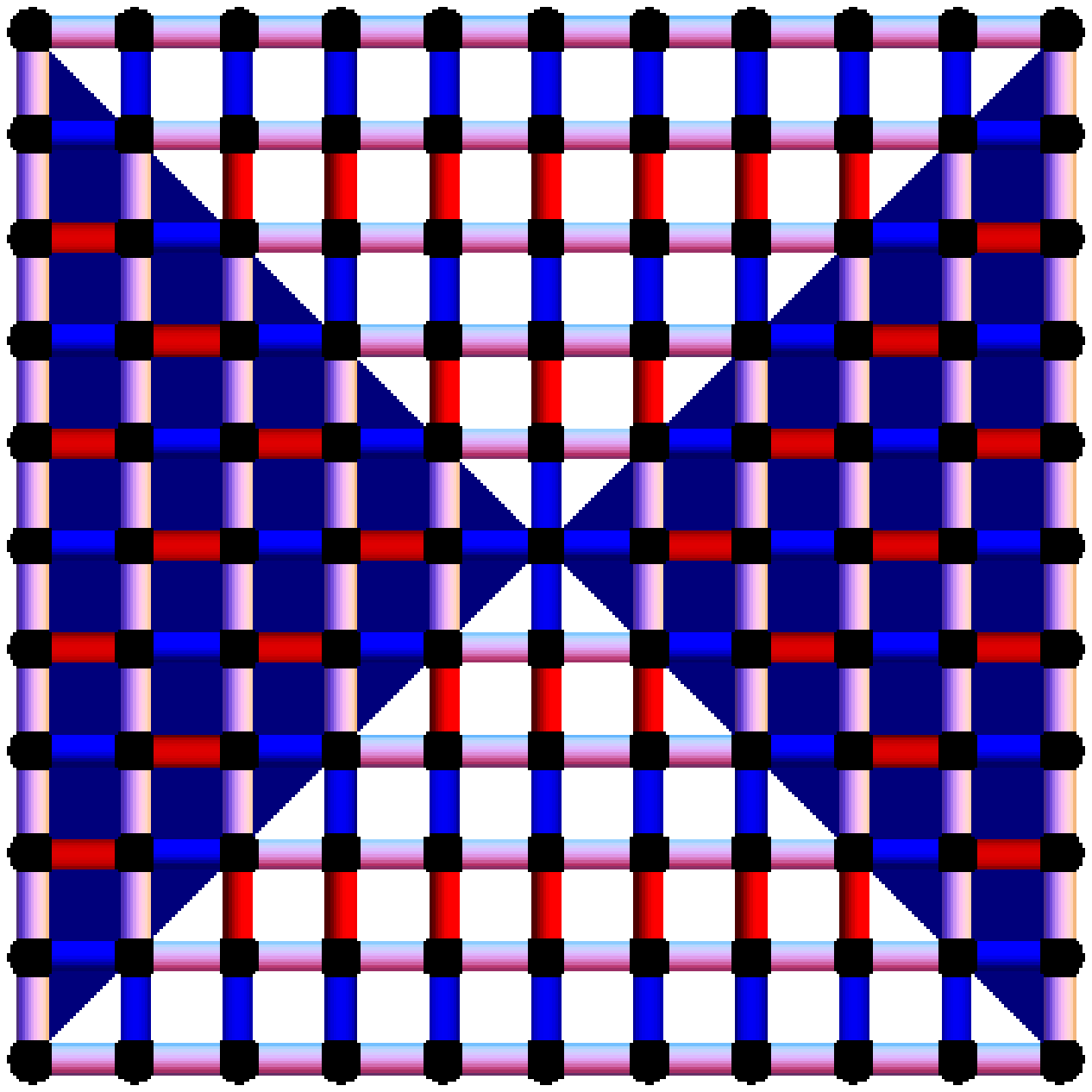}
\end{center}
\caption{Contributions to the hopping integrals 
from  a) Anti-vortex in $\beta$ term and
b) Anti-vortex in $m$ term. The color coding in a) is
blue for $J_-$, red for $J_+$, green for $-J_+$ and 
white for $-J_-$.  The color coding in b) is 
red for $m$ and 
blue for $-m$ (note that not all links have a contribution
to the hopping in case b)).  We have suppressed the radial dependence
of the hopping parameters for simplicity.  The central site is an A site.}
\label{fig:vortex}
\end{figure}

The results we obtained regarding the zero modes in Sec.~\ref{sec:zero} are
for a continuum theory.  Previous studies\cite{Seradjeh,Chamon2} have found 
that zero modes that are predicted from a continuum calculation are present
in the spectrum of appropriate tight-binding models on a lattice.
Given the complicated form of the vortex configuration we 
considered here, we check that the results we have obtained in the 
continuum limit carry over to the lattice.  To do this we 
considered the lattice model introduced in Ref.~\onlinecite{Kennett} 
and introduced a $Z_4$ vortex (anti-vortex) with vorticity 2 
in $\beta$ and a $Z_4$ vortex (anti-vortex) in $m$, as 
illustrated in Fig.~\ref{fig:vortex} for antivortices in $\beta$
and $m$ centered on an A site.

\begin{figure}[htb]
\includegraphics[width=8cm]{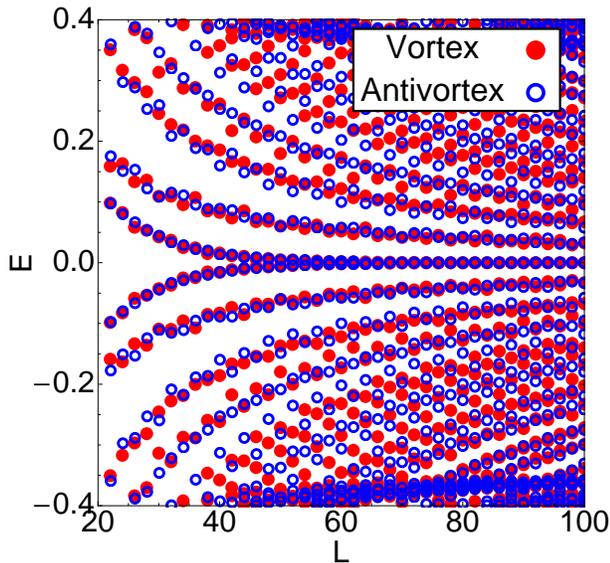}
\caption{Energy spectrum as a function of system size for vortex
and anti-vortex with $\beta_0 = 0.25$, $m_0 = 0.25$ and topological 
defect centered on site A.}
\label{fig:spectrum}
\end{figure}

We diagonalized the Hamiltonian on an $L \times L$ lattice 
and considered topological defects centered on A, B, C and D sites.  
We chose a vortex (anti-vortex) in both $\beta$ and $m$ with
step-function spatial profile $f(r) \propto \theta(r - r_0)$,
with $r_0 = 3$ lattice spacings.
Our results were qualitatively similar with topological defects centred
on different sites, and we mainly display data for configurations of the type 
shown in Fig.~\ref{fig:vortex} in which there is a topological defect
centered on an A site.
The spectrum as a function of system size is shown in 
Fig.~\ref{fig:spectrum} for $m_0 = 0.25$, $\beta_0 = 0.25$ for a topological
defect centered on an A site.  
We see that there are two states
which converge to zero energy, and a series of non-zero energy
states in the gap -- the continuum levels are visible at the top and 
bottom of the energy range.  We can also see that there 
are small differences in the spectrum when there are vortex
and anti-vortex configurations of the hopping.

\begin{figure}[htb]
\includegraphics[width=6.5cm,angle=270]{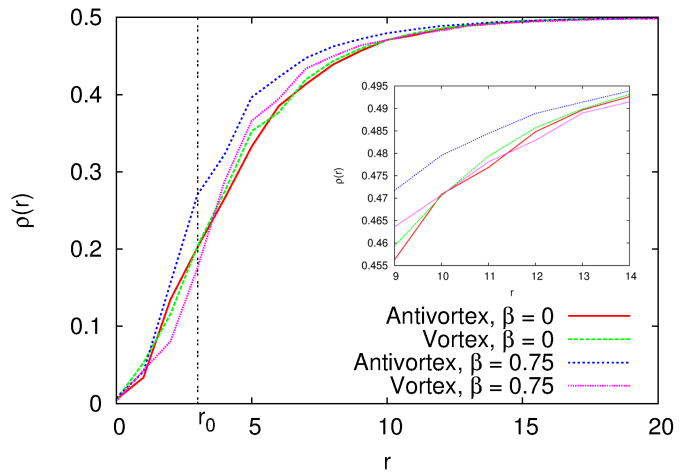}
\caption{Integrated charge density near the centre of a 102$\times$102 size system 
showing the spatial profile of the vortex and anti-vortex solutions 
with $m_0 = 0.25$ and $\beta_0 = 0$, and the spatial profile of the 
vortex and anti-vortex solutions when  $m_0 = 0.25$ and $\beta_0 = 0.75$, averaged
over defects located on A, B, C, and D sites.}
\label{fig:density}
\end{figure}

Even though there is only one vortex (anti-vortex) on the lattice, there are two
zero modes in the numerical spectrum as is expected for a finite size system 
with open boundary conditions.\cite{Chamon2}  
The probability density of these
states is localized at the centre of the lattice on the vortex (anti-vortex)
and at the edge of the system.  The integrated charge density 
averaged over all four possible locations of the topological defect (A, B, C or D)
for the zero energy modes is illustrated in Fig.~\ref{fig:density}.
We compare the cases  $m_0 = 0.25$ and $\beta_0 = 0$ and $m_0 = 0.25$ and $\beta_0 = 0.75$,
and it is quite evident that there is precisely half a charge localized around the centre of the 
system.  The profiles also are in qualitative agreement with the spatial dependences derived in 
Sec.~\ref{sec:zero}.  The antivortex solution with $\beta_0 = 0.75$ is 
considerably more localised than the vortex solution with the same $\beta_0$ and the solution when $\beta_0 = 0$, 
both at small $r$ and at larger $r \gtrsim 3 r_0$ (see the inset to Fig.~\ref{fig:density}).

\begin{figure}[htb]
\includegraphics[width=9cm]{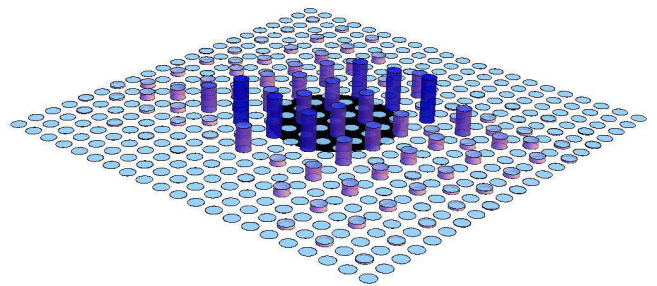}
\caption{Charge density near the centre of a 102$\times$102 size system centered on an A site
showing the spatial profile of the vortex  solutions when  $m_0 = 0.25$ and $\beta_0 = 0.75$ (the 
core of the vortex is shaded).
Note that there is support for the state only on A and D sites.}
\label{fig:wavefunction}
\end{figure}

A feature of our analytical results is that the vortex and antivortex solutions have
support on different sublattices.  This is also borne out in our numerical results.  We show
the charge density for a vortex with $m_0 = 0.25$ and $\beta_0 = 0.75$ in Fig.~\ref{fig:wavefunction} 
in which there is support for the state only on A and D sites, as deduced in Sec.~\ref{sec:zero}.
We also confirmed that the state in the presence of an anti-vortex only has support on B and C 
sites.

\section{Discussion and Conclusions}
\label{sec:disc}
In this paper we have studied the zero modes of birefringent Dirac
fermions arising when there is an unusual vortex configuration which involves
the kinetic energy as well as a mass term.   This situation 
differs from the usual one in which there is fractionalization of non-interacting
fermions in one or two dimensions due to a toplogical defect in some mass term.\cite{footnote2}
In order to have a
normalizable single-valued solution, there is a constraint on the allowed
vorticity of the vortex (anti-vortex) in $\beta$ in that it winds one more
time (in the same sense) as the vortex (anti-vortex) in $m$.

The effect of the topological defect in $\beta$ that we consider 
is to differentiate the spatial profiles of the zero modes associated
with vortex and anti-vortex zero modes.  The presence of
the toplogical defect in the kinetic energy in addition to the mass leads
to the unusual situation in which the length scale associated with the
localized state at a vortex differs from the length scale associated
with a state centered on an anti-vortex: the ratio of the two characteristic 
length scales is $(1+\beta_0)/(1-\beta_0)$
and so can be made arbitrarily large.  The role of the vortex and the
anti-vortex can be interchanged by changing the sign of $\beta$.  As we emphasised
in Sec.~\ref{sec:zero} the introduction of $\beta$ breaks the chiral symmetry that
relates the vortex and antivortex Hamiltonians, allowing the possibility of
differing zero mode solutions.
It should be noted that the topological defect in $\beta$ does not
affect the fractionalization associated with the topological defects,
only the relevant lengthscales.  We demonstrated this feature qualitatively
through diagonalizing a lattice model and comparing the integrated charge
density for the cases of $\beta_0 = 0$ and finite $\beta_0$.   

The results we have obtained here have interest in a broader context
than the particular problem we studied.  Our work gives an 
example of a physical property that systems 
which exhibit birefringent massless fermionic 
excitations\cite{Watanabe,Lan1,Lan2,Goldman,Moessner} 
can have that are  unavailable for regular Dirac fermions
 (an other such example is birefringent Klein tunnelling\cite{Lan1}).  
 The unusual occurrence
of zero modes with different lengthscales (but not with associated fractionalization)
was noted by two of us in Ref.~\onlinecite{Kennett} in the context of zero modes
for birefringent Dirac fermions in the presence of a domain wall in a $\gamma_0$ mass term,
and may be a generic feature of zero modes in birefringent Dirac systems.  A situation
in which there are two zero modes (hence there is no fractionalization)
associated with a topological defect, but each has a 
differing lengthscale was also recently discussed by one of us in the context of
graphene.\cite{Bitan}

\section{Acknowledgements}

The authors acknowledge helpful discussions with  Claudio
Chamon, Igor Herbut, Nazanin Komeilizadeh, Chi-Ken Lu and Joseph Thywissen.
We acknowledge support from NSERC.

\begin{appendix}
\section{Dispersion of birefringent Dirac fermions in the presence of a $i\gamma_0\gamma_3$ mass term.}
\label{app:gap}
In order for there to be localized zero modes in the presence of a topological defect, we need to be certain
that the additional term in the Hamiltonian leads to a gap in the 
spectrum in the absence of such a defect.  To confirm this, we calculate
the spectrum for the following Hamiltonian:
\begin{eqnarray}
H_k = i\gamma_0\gamma_1 k_x + i\gamma_0\gamma_2 k_y - \beta \gamma_3 k_x - \beta \gamma_5 k_y - m i\gamma_0\gamma_3. \nonumber \\
\end{eqnarray}
After a short calculation one may determine that the eigenvalues are
$$ \epsilon_k = \pm\sqrt{\left(1+\beta^2\right)|k|^2 + m^2 \pm 2\beta|k|^2 \sqrt{1 + \frac{m^2 k_y^2}{|k|^4}}}.$$
Note that there is always a gap of $2m$ at $k=0$, and that the minimum energy can occur at a finite value of 
$\bvec{k}$.  For appropriate choices of $\beta$ and $m$ there can be a gap, and the choices we make for 
our numerical calculations in Sec.~\ref{sec:numerics} are such that a gap always exists.  
An analogous calculation with an $i\gamma_0\gamma_5$ mass will yield a result such as this 
with $mk_y$ replaced by $mk_x$.

\end{appendix}


\begin{thebibliography}{99}
\bibitem{footnote} Note that in
Ref.~\onlinecite{Kennett} a Minkowski rather than Euclidean
metric was used, which corresponds  to slightly modified expressions
for  $\gamma_{1,2,3,5}$.
\bibitem{footnote2} There has been some study of momentum space vortices in the context
of pseudospin magnetism in graphene,\cite{Min} but there has
been no discussion of fractionalization in such a case.
\end{thebibliography}
\end{document}